# Study of new physics effects in lepton flavour violating $B \to K_2^*(1430)l_1l_2$ decays

**S. Biswas[1], M. Mandal[2], S. Mahata[3] and S. Sahoo[4]**

[1,2,3,4]Department of Physics, National Institute of Technology Durgapur
Durgapur-713209, West Bengal, India

[1]E-mail: getswagata92@gmail.com, [4]E-mail: sukadevsahoo@yahoo.com

**Abstract**

Lepton flavour violation (LFV) is one of the most trending topics to probe new physics (NP). The powerful accelerators have enhanced their intensities to observe the LFV decays very precisely. In this situation, the theorists are also interested to study these decays in various NP models and in model independent way to get precise results. Motivated by these results we have studied $B \to K_2^*(1430)l_1l_2$ decays in non-universal $Z'$ model. Here, we have structured the two-fold angular distribution of the decays in terms of transversity amplitudes and the transversity amplitudes are formed with NP Wilson coefficients. The variation of the branching ratios and forward backward asymmetries show the sensitivity of NP. The observables calculated in this work are very interesting and might provide a new way towards NP.

## I. Introduction:

The discovery of Higgs boson completed the family picture of the standard model (SM) of particle physics. The SM has successfully described most of the phenomena of nature at the energies near to electroweak scale. But now it is accepted that the SM needs to be modified to explain the new physics (NP) phenomenologies such as origin of dark matter, neutrino oscillations, the well-known flavour puzzle, etc. The experimental status of B anomalies is recorded by the B factories as well as the LHCb. Nowadays, the lepton flavour universality test is done by measuring $R_{D,D^*}$. The observable $R_{D^*}$ is measured at Belle [1, 2], BaBar [3] and LHCb [4]. The recent measurements of Belle [5] are: $R_D = 0.307 \pm 0.37 \pm 0.016$ and $R_{D^*} = 0.283 \pm 0.018 \pm 0.14$. These results are greater than the SM predictions calculated in ref. [6] and ref. [7] by $2.3\sigma$ and $3.4\sigma$ deviations respectively. The LHCb [4] reported their preliminary result of $R_D$ and $R_{D^*}$ as $R_D^{LHCb2022} = 0.441 \pm 0.060 \pm 0.066$ and $R_{D^*}^{LHCb2022} = 0.281 \pm 0.018 \pm 0.024$. The world averages of the $R_{D^{(*)}}$ measurements are [8]: $R_D = 0.358 \pm 0.025 \pm 0.012$ and $R_{D^*} = 0.285 \pm 0.010 \pm 0.008$. Though the latest measurements of $R_K$ and $R_{K^*}$ are found to be in agreement with the SM value in the bin range $0.045 \leq q^2 \leq 1.1$ and $1.1 \leq q^2 \leq 6.0$ [9-12] (where $q^2$ is the momentum transfer term), there are lots of footprints for the existence of NP.

Another recent topic to test the SM is the lepton flavour violation (LFV). Study on LFV decays was started with the discovery of muon particle as a separate one at early 1940's. The evidence of quark flavour violation is very prompt in the colliders whereas the lepton flavour violation is noticeable but is not established experimentally. The neutrino oscillation process indicates the fact of neutrino flavour violation. As all the fermions other than leptons in the SM of particle physics show flavour violation, we can expect that the charged leptons could mix. It is already known that if the neutrinos have masses and get mixed then the mixing can be transferred to charged leptons via $\nu_l - e - W$ coupling. The experiments have proclaimed that the neutrinos are massive and they mix, so it is obvious to exist mixings among the charged leptons. The SM predicts neutrinos to be massless which contradicts the accelerators. To include massive neutrinos into the theory we need to extend the SM. The neutrino mass is very tiny, approximately smaller than 1eV and this propels the neutrino Yukawa coupling to be of $10^{-12}$. The LFV decays such as $B^0 \to \tau^\pm \mu^\mp$ and $B_s^0 \to \tau^\pm \mu^\mp$ are extremely suppressed in the SM with expected branching ratios of the order of $10^{-54}$ [13] whereas the LHCb [14] has reported the upper limits for the branching ratios at 90% confidence level as $\mathcal{B}(B^0 \to \tau^\pm \mu^\mp) < 1.4 \times 10^{-5}$ and $\mathcal{B}(B_s^0 \to \tau^\pm \mu^\mp) < 4.2 \times 10^{-5}$. Recently, the Belle Collaboration [15] has set upper limit on branching ratio at 90% confidence level as $\mathcal{B}(B_s^0 \to \mu^\mp \tau^\pm) < 7.3 \times 10^{-4}$.

Here, we have chosen the $B \to K_2^*$ channel because the colliders have provided better understanding of the radiative decays in inclusive level with the $B \to K_2^* \gamma$ decay. On the other hand, the SM branching ratio value for $B \to K_2^*(1430)ll$ shows that this decay channel can provide an independent test of the SM [16, 17]. Therefore, we have selected this channel in LFV mode to probe the NP. We hope the precise measurement of $B \to K_2^*(1430)l_1 l_2$ decays would be possible with the increasing of the sensitivity of the current experiments in future.

There are several theoretical attempts to demonstrate the experimental tensions which are discussed above. Though there are only experimental bounds exist on the LFV decays, various NP models have explained them. These decays are studied with the effect of FCNC mediated $Z$ boson [18, 19], in non-universal $Z'$ model [20-22], in leptoquark model [23-26], in MSSM [27-29] and other NP models [30] and also in model independent way [31]. Previously, Biswas et al. have studied the LFV $\Lambda_b$ decays [32], $B \to K^* l_1 l_2$ and $B_s \to \varphi l_1 l_2$ decays [33] in non-universal $Z'$ model where $l_1$ and $l_2$ are charged leptons of different flavours. Here, our work is based on the same NP model mentioned previously. In this paper, we have studied $B \to K_2^*(1430) l_1 l_2$ decays in non-universal $Z'$ model. Structuring the two-fold decay distribution of the decays we have investigated the sensitivity of NP on various observables with the contribution of vector, axial-vector NP operators. The non-universal $Z'$ model explains the NP contribution at tree level through $Z'$-mediated flavour changing $b \to s$ transitions. The $Z'$ boson associates to the quark sector as well as to the leptonic sector. The $Z'$ boson is not observed at the colliders till now but its mass is constrained differently in different in grand unified theories (GUTs). Various experiments and detectors have also restricted the upper and lower bounds of the mass of $Z'$ boson. The model-dependent lower bound for $Z'$ mass is set at 500 GeV by different accelerators [34-36]. Sahoo et al. have predicted the range of $Z'$ mass as $1352 - 1665$ GeV from $B_q - \overline{B_q}$ mixing [37]. Several other studies have also set the mass limits which are discussed in the refs. [38-41]. The ATLAS collaboration has constrained the lower bound of $Z'$ mass for $Z'_\psi$ and $Z'_{SSM}$ as 4.5 TeV and 5.1 TeV respectively at 95% C. L. [42]. Recently [43], the mass difference of $B_s$ meson is studied in extended standard model and the upper limit of $Z'$ mass is constrained as 9 TeV. In this paper, we have taken the $Z'$ mass in TeV range.

This paper is arranged as follows: The effective Hamiltonian for LFV decays is discussed in Sec. II. The details of the decay and the description of the observables are illustrated in Sec. III. In Sec. IV, the contribution of non-universal $Z'$ model is incorporated in the decays by modifying the Wilson coefficients. We have presented the numerical analysis in Sec. V. And finally, we have concluded the findings in Sec. VI.

## II. Effective Hamiltonian:

In this section, we structure the effective Hamiltonian for the lepton flavour violating $b \to s l_1 l_2$ transition. The leptons $l_1$ and $l_2$ are considered of the same flavour $l$ in the SM but in BSM physics the NP particle $Z'$ will couple differently with leptons of different families. The structured Hamiltonian is represented as follows [21, 27, 30]

$$\mathcal{H}^{eff} = -\frac{G_F \alpha}{2\sqrt{2}\pi} V_{tb} V_{ts}^* \sum_{r=9,10} C_r^{NP} O_r + h.c., \quad (1)$$

where $G_F$ represents the Fermi coupling constant, $\alpha$ represents the electromagnetic coupling constant. The $C_r^{NP}$ parts are the Wilson Coefficients containing NP contributions. The CKM matrix elements $V_{tb} V_{ts}^*$ are introduced in the Hamiltonian due to the virtual effects induced by $t\bar{t}$ loops. It is to be noted that the LFV decays occur at tree level in this $Z'$ model; therefore, the NP should contribute in the fashion where $t\bar{t}$ loops get cancelled. Moreover, there is an electromagnetic operator $O_7$ in the SM for $b \to sll$ transition. Non-universal $Z'$ model is basically sensitive to the semileptonic current operators $O_9$ and $O_{10}$ involving NP contributions in $C_9^{NP}$ and $C_{10}^{NP}$ [21, 26, 44]. Here,

$$\begin{aligned} O_9 &= [\bar{s}\gamma_\mu(1-\gamma_5)b][\bar{l}_1 \gamma^\mu l_2], \\ O_{10} &= [\bar{s}\gamma_\mu(1-\gamma_5)b][\bar{l}_1 \gamma^\mu \gamma_5 l_2]. \end{aligned} \quad (2)$$

## III. The $B \to K_2^*(1430) l_1 l_2$ decays

The $B \to K_2^*(1430) l_1 l_2$ decay can be structured in terms of following kinematic variables: (i) $\theta_l$, the angle made by $l_1$ lepton with z axis in the dilepton rest frame, (ii) $q^2 (= (p-k)^2)$, the four momentum of dilepton pair (where $p$ and $k$ are the four momentum of $B$ and $K_2^*$ mesons respectively). The two fold decay differential branching ratio can be described as [32, 45, 46]

$$\frac{d^2 \mathcal{B}}{dq^2 d\cos\theta_l} = A(q^2) + B(q^2)\cos\theta_l + C(q^2)\cos^2\theta_l. \quad (3)$$

Here, the terms $A(q^2)$, $B(q^2)$ and $C(q^2)$ can be written as follows

$$A(q^2) = \frac{3}{4}\left\{\frac{1}{4}\left[\left(1+\frac{m_+^2}{q^2}\right)\beta_-^2 + \left(1+\frac{m_-^2}{q^2}\right)\beta_+^2\right]\left(|A_L^\parallel|^2 + |A_L^\perp|^2 + (L\leftrightarrow R)\right)\right.$$
$$+ \frac{1}{2}(\beta_+^2 + \beta_-^2)(|A_L^0|^2 + |A_R^0|^2)$$
$$+ \frac{4m_1 m_2}{q^2}Re\left[A_R^0 A_L^{0*} + A_R^\parallel A_L^{\parallel *} + A_R^\perp A_L^{\perp *} - A_L^t A_R^{t*}\right]$$
$$\left.+ \frac{1}{2}(\beta_+^2 + \beta_-^2 - 2\beta_-^2\beta_+^2)(|A_L^t|^2 + |A_R^t|^2)\right\}, \tag{4}$$

$$B(q^2) = \frac{3}{2}\beta_-\beta_+\left\{Re[A_L^{\perp *}A_L^\parallel - (L\leftrightarrow R)] + \frac{m_+ m_-}{q^2}Re[A_L^{0*}A_L^t + (L\leftrightarrow R)]\right\}, \tag{5}$$

$$C(q^2) = \frac{3}{8}\beta_+^2\beta_-^2\left\{\left(\left|A_L^\parallel\right|^2 + |A_L^\perp|^2 - 2|A_L^0|^2\right) + (L\leftrightarrow R)\right\}, \tag{6}$$

where $m_\pm = (m_1 \pm m_2)$, $\beta_\pm = \sqrt{1 - \frac{(m_1 \pm m_2)^2}{q^2}}$ and $m_1, m_2$ are the masses of leptons $l_1$ and $l_2$ respectively. The angular coefficients of the differential branching fraction of eq. (3) are expressed in terms of the transversity amplitudes which are described in the Appendix A. The transversity amplitudes are structured in terms of the Wilson coefficients and the form factors. Basically, the short distance physics are incorporated at the NP Wilson coefficients and the long distance physics are inserted via $B \to K_2^*$ hadronic elements.

The hadronic matrix elements for $B \to K_2^*$ transition are parameterized in terms of four form factors $V(q^2)$ and $A_{0,1,2}(q^2)$. The matrix elements for vector and axial vector currents are structured as follows

$$\langle K_2^*(k,\epsilon^*)|\bar{s}\gamma^\mu b|\bar{B}(p)\rangle = -\frac{2V(q^2)}{m_B + m_{K_2^*}}\epsilon^{\mu\nu\rho\sigma}\epsilon^*_{T\nu}p_\rho k_\sigma, \tag{7}$$

$$\langle K_2^*(k,\epsilon^*)|\bar{s}\gamma^\mu\gamma_5 b|\bar{B}(p)\rangle$$
$$= 2im_{K_2^*}A_0(q^2)\frac{\epsilon_T^*\cdot q}{q^2}q^\mu + i(m_B + m_{K_2^*})A_1(q^2)\left[\epsilon_T^{*\mu} - \frac{\epsilon_T^*\cdot q}{q^2}q^\mu\right]$$
$$- iA_2(q^2)\frac{\epsilon_T^*\cdot q}{(m_B + m_{K_2^*})}\left[(p+k)^\mu - \frac{(m_B^2 - m_{K_2^*}^2)}{q^2}q^\mu\right]. \tag{8}$$

The form factors used in this calculation are derived using the light cone QCD sum rule. We have used the updated values of the form factors from the reference [47]. The numerical values of the form factors are collected in the Appendix B. The $K_2^*$ meson is the higher excited state of $K^*$ meson having spin-2. The details of the polarization vector of $K_2^*$ is discussed in the Appendix C. In our analysis, we have structured the leptonic helicity amplitudes which are discussed briefly in the Appendix D.

With all these structures, we have calculated the differential decay rate by integrating over the angle $\theta_l$ as

$$\frac{d\mathcal{B}}{dq^2} = 2\left[A(q^2) + \frac{C(q^2)}{3}\right]. \tag{9}$$

To probe NP in LFV decays we have structured another powerful tool, the forward-backward asymmetry. It is defined as

$$A_{FB}(q^2) = \frac{\int_0^1 \frac{d^2\mathcal{B}}{dq^2 d\cos\theta_l} d\cos\theta_l - \int_{-1}^0 \frac{d^2\mathcal{B}}{dq^2 d\cos\theta_l} d\cos\theta_l}{\int_0^1 \frac{d^2\mathcal{B}}{dq^2 d\cos\theta_l} d\cos\theta_l + \int_{-1}^0 \frac{d^2\mathcal{B}}{dq^2 d\cos\theta_l} d\cos\theta_l}. \tag{10}$$

The observable can be expressed as

$$A_{FB}(q^2) = \frac{B(q^2)}{2\left[A(q^2) + \frac{C(q^2)}{3}\right]}. \tag{11}$$

## IV. The contribution of $Z'$ boson in $B \to K_2^*(1430)l_1 l_2$ decays

The presence of $Z'$ boson can reform the effective Hamiltonian of $b \to s$ transitions to provide an appreciable deviation from the SM values and to explain the collider results. An extra $U(1)'$ gauge group is introduced with the SM gauge group [48, 49] and the $Z'$ boson is evolved through spontaneous symmetry breaking process. This new massive boson couples to quarks as well as the lepton pair and can explain the FCNC transitions at the tree level. According to refs. [50-52] $Z'$ boson associates with the third generation of quark differently from the other two generations and show similar behaviour for the lepton families also. In this paper, we have taken different family non-universal $Z'$ couplings for different lepton families in the model. These couplings are diagonal and non-universal in nature.

The current can be represented in NP as

$$J_\mu = \sum_{i,j} \bar{\psi}_j \gamma_\mu \left[\epsilon_{\psi_{L_{ij}}} P_L + \epsilon_{\psi_{R_{ij}}} P_R\right] \psi_i. \tag{12}$$

Here, this sum is applied for all fermions $\psi_{i,j}$ and $\epsilon_{\psi_{R,L_{ij}}}$ are the chiral couplings of the new gauge boson. The FCNCs are explained in both left-handed and right-handed sectors at the tree level in $Z'$ model. Therefore, we can represent $B_{ij}^{\psi_L} \equiv \left(V_L^\psi \epsilon_{\psi_L} V_L^{\psi\dagger}\right)_{ij}$ and $B_{ij}^{\psi_R} \equiv \left(V_R^\psi \epsilon_{\psi_R} V_R^{\psi\dagger}\right)_{ij}$. The NP coupling is introduced as

$$\mathcal{L}_{FCNC}^{Z'} = -g'\left(B_{sb}^L \bar{s}_L \gamma_\mu b_L + B_{sb}^R \bar{s}_R \gamma_\mu b_R\right) Z'^\mu + h.c., \tag{13}$$

where $g'$ is the new gauge coupling linked with the extra $U(1)'$ group and the effective Hamiltonian becomes,

$$H_{eff}^{Z'} = \frac{8G_F}{\sqrt{2}} \left(\rho_{sb}^L \bar{s}_L \gamma_\mu b_L + \rho_{sb}^R \bar{s}_R \gamma_\mu b_R\right) \left(\rho_{l_i l_j}^L \bar{l}_{j_L} \gamma_\mu l_{i_L} + \rho_{l_i l_j}^R \bar{l}_{j_R} \gamma_\mu l_{i_R}\right), \tag{14}$$

where $\rho_{l_i l_j}^{L,R} \equiv \frac{g' M_Z}{g M_{Z'}} B_{l_i l_j}^{L,R}$, $g$ and $g'$ are the gauge couplings of $Z$ and $Z'$ bosons (here, $g = \frac{e}{\sin\theta_W \cos\theta_W}$) respectively.

In this work, we have incorporated some simplifications regarding this NP model which are:

(i) we have ignored kinetic mixing because it represents the redefinition of unknown couplings,

(ii) we have also neglected the mixing of $Z - Z'$ [48, 53-56] for its very small value. The upper bound of mixing angle is found $10^{-3}$ by Bandyopadhyay et al. [40] and $10^{-4}$ by Bobovnikov et al. [57],

(iii) we have considered that there are no significant contributions of renormalization group (RG) evolution between $M_W$ and $M_{Z'}$ scales,

(iv) we accept the considerable contribution of the flavour-off-diagonal left-handed quark couplings in the flavour changing quark transition [58-62] in our investigation. The detail of this assumption is discussed in our previous work [32],

(v) here, the value of the term $\left|\frac{g'}{g}\right|$ is not fixed yet. As both $U(1)$ gauge groups, included in this $Z'$ model, are generated from the same origin of some GUT, we have taken $\left|\frac{g'}{g}\right| \sim 1$,

(vi) we have taken $\left|\frac{M_Z}{M_{Z'}}\right| \sim 0.1$ for $Z'$ of TeV-scale.

The LEP experiments have also recommended the $Z'$ existence with the identical couplings as the SM $Z$ boson. We can say that if $|\rho_{sb}^L| \sim |V_{tb} V_{ts}^*|$, then $B_{sb}^L$ will be $\mathcal{O}(10^{-3})$. Considering all above discussions, we can structure the effective Hamiltonian for the LFV $b \to s l_1 l_2$ transition as

$$H_{eff}^{Z'} = -\frac{2G_F}{\sqrt{2}\pi} V_{tb} V_{ts}^* \left[ \frac{B_{sb}^L B_{l_1 l_2}^L}{V_{tb} V_{ts}^*} (\bar{s}b)_{V-A} (\bar{l}_1 l_2)_{V-A} + \frac{B_{sb}^L B_{l_1 l_2}^R}{V_{tb} V_{ts}^*} (\bar{s}b)_{V-A} (\bar{l}_1 l_2)_{V+A} \right] + h.c, (15)$$

where $B_{sb}^L$ is the left-handed coupling of $Z'$ boson with quarks, $B_{l_1 l_2}^L$ and $B_{l_1 l_2}^R$ are the left-handed and right-handed couplings with the leptons respectively. The NP quark coupling consists of a NP weak phase term, which is related as $B_{sb}^L = |B_{sb}^L| e^{-i\varphi_s^l}$. The Wilson coefficients can be structured including the NP terms as

$$C_9^{NP} = \frac{4\pi B_{sb}^L}{\alpha V_{tb} V_{ts}^*} \left( B_{l_1 l_2}^L + B_{l_1 l_2}^R \right),$$
$$C_{10}^{NP} = \frac{4\pi B_{sb}^L}{\alpha V_{tb} V_{ts}^*} \left( B_{l_1 l_2}^L - B_{l_1 l_2}^R \right). \quad (16)$$

Including the NP parts through the modified Wilson coefficients, we have studied the observables defined in eqs. (9) and (11) in the non-universal $Z'$ model.

## V. Numerical Analysis

In this work, we have studied differential branching ratios and forward backward asymmetries for the LFV decays $B \to K_2^*(1430)\mu\tau$ and $B \to K_2^*(1430)e\mu$ in the framework of non-universal $Z'$ model. We have mainly modified the terms $C_9^{NP}$ and $C_{10}^{NP}$ with the NP coupling terms which are constrained from $B_s - \bar{B}_s$ mixing data of UTfit Collaboration [63-69] and various inclusive and exclusive decays. The numerical values of the $b - s - Z'$ couplings and the NP weak phases $\varphi_s^l$ are tabulated below in Table-1. The first two scenarios are taken from refs. [44, 70]. The third scenario is constrained from the recent values of the mass differences updated in ref. [43]. Using the recent updated values of mass differences, we have constrained NP terms at our previous work [71]. Other input parameters are recorded in Appendix E [72].

**Table-1: Numerical values of coupling parameters**

| Scenarios | $\|B_{sb}\| \times 10^{-3}$ | $\varphi_s^l$ (in degree) |
|---|---|---|
| $S_1$ | $(1.09 \pm 0.22)$ | $(-72 \pm 7)°$ |
| $S_2$ | $(2.20 \pm 0.15)$ | $(-82 \pm 4)°$ |
| $S_3$ | $0 \leq \|B_{sb}^L\| \leq 1.539 \times 10^{-3}$ | For $0° \leq \varphi_s^L \leq 180°$ |

To increase as well as to examine the sensitivity of NP in our investigation we have posed the maximum values of the NP couplings in the observables. With the help of Table-1 we have selected three sets of $Z'$ couplings as below.

<u>For scenario 1 ($S_1$):</u> $|B_{sb}| = (1.31 \times 10^{-3})$ and $\varphi_s^l = -65°$.

<u>For scenario 2 ($S_2$):</u> $|B_{sb}| = (2.35 \times 10^{-3})$ and $\varphi_s^l = -78°$.

<u>For scenario 3 ($S_3$):</u> $|B_{sb}| = (1.539 \times 10^{-3})$ and $\varphi_s^l = 180°$.

Here, in our work we have composed the leptonic couplings as $S_{l_1 l_2} = \left(B_{l_1 l_2}^L + B_{l_1 l_2}^R\right)$, and $D_{l_1 l_2} = \left(B_{l_1 l_2}^L - B_{l_1 l_2}^R\right)$. The maximally allowed region for leptonic couplings are found in the ref. [32] and these values are: $S_{\mu e} = 0.0079$, $D_{\mu e} = -0.0079$ and $S_{\tau \mu} = 0.11$, $D_{\tau \mu} = -0.11$. Another consideration we have adopted in this work is $C_9^{NP} = -C_{10}^{NP}$ because the $C_9^{NP} = C_{10}^{NP}$ scenario is very weak to produce effective results (which can be confirmed with the results of the refs. [73, 74]). On the other hand, the $C_9^{NP} = -C_{10}^{NP}$ scenario provides lots of fruitful results under the philosophy of both model and model-independent strategies. Some of these best results are presented in the refs. [26, 73-78].

With all these numerical data and theoretical considerations, we have varied the differential branching ratio within allowed kinematic region of $q^2$ in Fig. 1a and Fig. 1b for $B \to K_2^* \mu\tau$ and $B \to K_2^* e\mu$ channels respectively. Here, all the plots are drawn using the central values of the form factors and other input parameters. In the Fig. 1, the magenta line represents the 1$^{st}$ scenario, the red line represents the 2$^{nd}$ scenario and the blue line represents the 3$^{rd}$ scenario. We observe that the branching ratio has greater value on mid $q^2$ region for $B \to K_2^* \mu\tau$ transition and on low $q^2$ region for $B \to K_2^* e\mu$ transition. There is a difference between the natures of these differential branching ratios due to the lighter mass of electron. But both the decay channels have more branching value for higher contribution of NP which confirms the responsiveness of the non-universal $Z'$ model. Another conclusion which can be drawn here is that the maximum attained value of the observable for 2$^{nd}$ scenario is large in comparison to the other scenarios. Integrating over the whole $q^2$ phase space we have estimated the average values of branching ratios for these decays which are shown in Table-2. Here, we have presented the maximum allowed values of the branching ratios considering the maximum contribution of NP couplings.

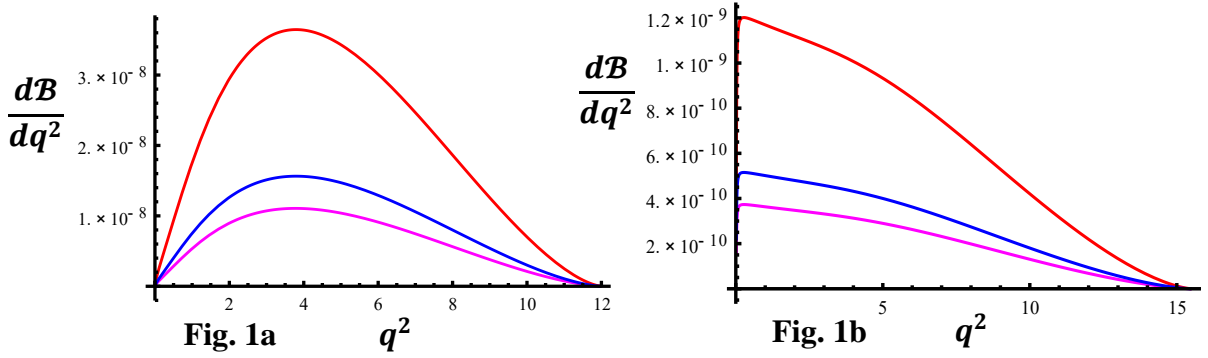

**Fig. 1: Variation of differential branching ratio within allowed kinematic region of $q^2$ using the bound of NP couplings for (a) $B \to K_2^* \mu\tau$, (b) $B \to K_2^* e\mu$.**

**Table-2: Predicted values of differential branching ratios for LFV $B \to K_2^* \mu\tau$ and $B \to K_2^* e\mu$ decays in 1$^{st}$, 2$^{nd}$ and 3$^{rd}$ scenarios**

| Kinematic region ($q^2$) (in GeV$^2$) | Differential branching ratio value | | |
|---|---|---|---|
| | For $B \to K_2^* \mu\tau$ mode | | |
| | 1$^{st}$ Scenario ($S_1$) | 2$^{nd}$ Scenario ($S_2$) | 3$^{rd}$ Scenario ($S_3$) |
| In $q^2 = 2$ | $(9.20 \pm 0.24) \times 10^{-9}$ | $(3.28 \pm 0.32) \times 10^{-8}$ | $(1.47 \pm 0.20) \times 10^{-8}$ |
| In $q^2 = 6$ | $(9.35 \pm 1.52) \times 10^{-9}$ | $(3.46 \pm 1.63) \times 10^{-8}$ | $(1.51 \pm 0.58) \times 10^{-8}$ |
| In $q^2 = 10$ | $(2.34 \pm 1.21) \times 10^{-9}$ | $(7.59 \pm 3.25) \times 10^{-9}$ | $(3.24 \pm 2.93) \times 10^{-9}$ |
| | For $B \to K_2^* e\mu$ mode | | |
| | 1$^{st}$ Scenario ($S_1$) | 2$^{nd}$ Scenario ($S_2$) | 3$^{rd}$ Scenario ($S_3$) |
| In $q^2 = 2$ | $(3.76 \pm 0.02) \times 10^{-10}$ | $(1.36 \pm 0.13) \times 10^{-9}$ | $(5.13 \pm 0.21) \times 10^{-10}$ |
| In $q^2 = 6$ | $(2.90 \pm 1.20) \times 10^{-10}$ | $(8.69 \pm 2.85) \times 10^{-10}$ | $(3.88 \pm 1.35) \times 10^{-10}$ |
| In $q^2 = 10$ | $(1.54 \pm 1.21) \times 10^{-10}$ | $(4.47 \pm 2.25) \times 10^{-10}$ | $(2.14 \pm 1.80) \times 10^{-10}$ |

To probe the $Z'$ contribution in the decay modes we have investigated the variation of forward backward asymmetries with respect to whole kinematic region. We have previously studied the observable in NP for LFV bottom baryonic and mesonic decays [32, 33]. We have studied the variation of forward-backward asymmetries in the $Z'$ model in the allowed $q^2$ region in Fig. 2a and Fig. 2b for $B \to K_2^* \mu\tau$ and $B \to K_2^* e\mu$ channels respectively. The colour description of the figures is similar to the previous ones. The change in the zero crossing positions interpret the sensitivity of $Z'$ boson on these LFV decays. According to the Figs. 2, the zero crossing point of $B \to K_2^* \mu\tau$ decay channel is at $q^2 = 7.585$ GeV$^2$ for 1$^{st}$ scenario, at $q^2 = 4.148$ GeV$^2$ for 2$^{nd}$ scenario and at $q^2 = 5.614$ GeV$^2$ for 3$^{rd}$ scenario. The observable is negative for low $q^2$ region and attained the maximum value at high $q^2$ region. Here, we observe that the zero crossings gradually shift towards the origin with the increment of NP contributions. Similar to differential branching ratios the forward backward asymmetries also have different nature for $B \to K_2^* e\mu$ decay channel. Here, the observable has very small negative value at low $q^2$ region and the variation is mainly positive throughout the kinematic region. The zero crossing point of $B \to K_2^* e\mu$ decay channel is at $q^2 = 0.748$ GeV$^2$ for 1$^{st}$ scenario, at $q^2 = 0.502$ GeV$^2$ for 2$^{nd}$ scenario and at $q^2 = 0.592$ GeV$^2$ for 3$^{rd}$ scenario. The zero crossing points have lesser value with more $Z'$ contributions in this $B \to K_2^* e\mu$ channel. The zero crossings are located clearly in Fig. 3a and Fig. 3b for $B \to K_2^* \mu\tau$ and $B \to K_2^* e\mu$ channels respectively. The values of the forward-backward asymmetries are calculated and represented in Table-3.

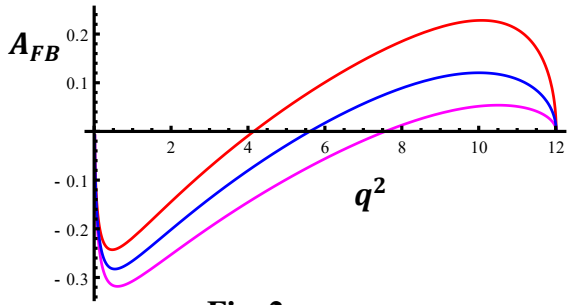 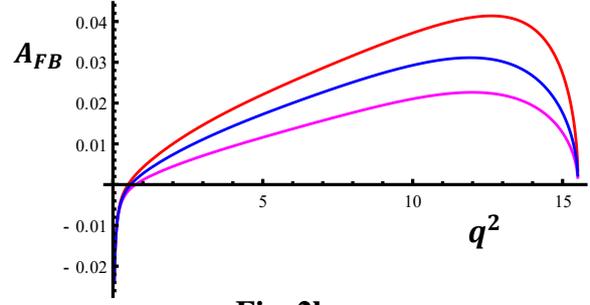

**Fig. 2a**  **Fig. 2b**

**Fig. 2: Variation of forward backward asymmetries within allowed kinematic region of $q^2$ using the bound of NP couplings for (a) $B \to K_2^* \mu\tau$, (b) $B \to K_2^* e\mu$.**

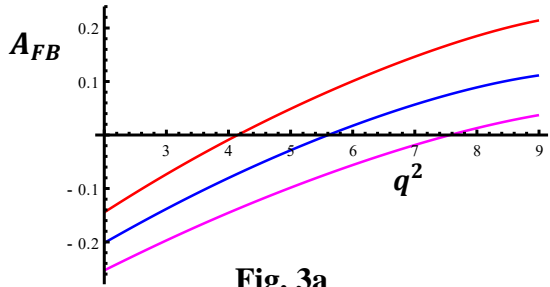 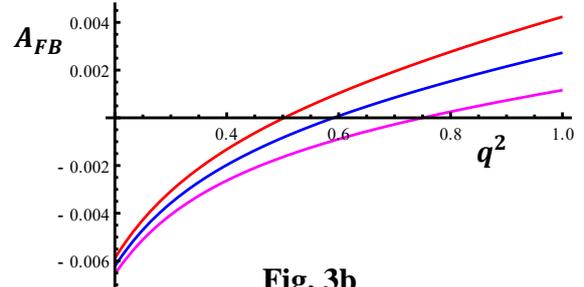

**Fig. 3a**  **Fig. 3b**

**Fig. 3: Variation of forward backward asymmetries within allowed kinematic region of $q^2$ using the bound of NP couplings for (a) $B \to K_2^* \mu\tau$, (b) $B \to K_2^* e\mu$ to locate zero crossing points.**

**Table-3: Predicted values of forward asymmetries for LFV $B \to K_2^* \mu\tau$ and $B \to K_2^* e\mu$ decays in 1st, 2nd and 3rd scenarios**

| Kinematic region ($q^2$) (in GeV²) | Forward backward asymmetry value | | |
|---|---|---|---|
| | For $B \to K_2^* \mu\tau$ mode | | |
| | 1st Scenario ($S_1$) | 2nd Scenario ($S_2$) | 3rd Scenario ($S_3$) |
| In $q^2 = 2$ | $-0.250 \pm 0.030$ | $-0.142 \pm 0.002$ | $-0.200 \pm 0.021$ |
| In $q^2 = 6$ | $-0.056 \pm 0.009$ | $0.105 \pm 0.010$ | $0.019 \pm 0.007$ |
| In $q^2 = 10$ | $0.054 \pm 1.206$ | $0.229 \pm 0.221$ | $0.125 \pm 0.111$ |
| | For $B \to K_2^* e\mu$ mode | | |
| | 1st Scenario ($S_1$) | 2nd Scenario ($S_2$) | 3rd Scenario ($S_3$) |
| In $q^2 = 2$ | $0.005 \pm 0.001$ | $0.011 \pm 0.002$ | $0.008 \pm 0.002$ |
| In $q^2 = 6$ | $0.014 \pm 0.010$ | $0.026 \pm 0.010$ | $0.021 \pm 0.010$ |
| In $q^2 = 10$ | $0.022 \pm 0.019$ | $0.038 \pm 0.028$ | $0.031 \pm 0.029$ |

## VI. Conclusion:

The lepton flavour non-universality in $b \to sl^+l^-$ transitions has been the tantalizing sector to probe NP over the years. Recently the accelerators have studied the LFV $h \to \mu\tau$ decays which becomes a clear cut hint for BSM physics to explore. If we have a look on PDG, we can see several updated measurements on LFV decays which enthuse the theoretical community to study the lepton flavour violation in various NP models as well as in model independent way. Previously we have found the NP LFV couplings [32] and studied the $b \to sl_1^- l_2^+$ decays in the refs. [32, 33]. In this work, we have chosen the excited state of $K^*$ meson with spin two. The lepton flavour conserving and violating decays of this $K_2^*$ meson are previously studied with effect of vector leptoquark [45, 79]. We have studied the LFV $B \to K_2^* \mu\tau$ and $B \to K_2^* e\mu$ channels in the non-universal $Z'$ model. Here, we have calculated the branching ratio values and forward backward asymmetries for these decays. The NP couplings which are constrained in the ref. [32] have provided fruitful results and have shown the sensitivity of the $Z'$ boson in the channel. The variation of the observables over the whole kinematic region shows that the 2nd scenario provides the maximum values of the observables which infers the clear cut signature of the NP. Here, the change in zero crossing positions with respect to three scenarios conveys the responsivity of NP in the decays. From the Fig. 3a and Fig. 3b, we see that the zero crossings approach the origin with the increment of contribution of non-universal $Z'$ boson. Therefore, it can be stated that the contribution of $Z'$ boson flips the value of $A_{FB}$ faster and it attains the maximum value at high $q^2$ regime. Another thing we observe from the Fig. 1, 2, 3 that the nature of the observables for $B \to K_2^* \mu\tau$ decay channel is quite different from $B \to K_2^* e\mu$ decay channel due to the difference in leptonic masses. This variance of character probes the NP on the LFV decays.

The authors in ref. [45] have studied the $B \to K_2^* \mu\tau$ decay channel and calculated the maximum value of branching ratio over whole kinematic region as $1.64 \times 10^{-7}$ in vector leptoquark model whereas the average values of the branching ratios are calculated as $(0.63 \times 10^{-8})$ for 1st scenario, $(2.09 \times 10^{-8})$ for 2nd scenario and $(0.89 \times 10^{-8})$ for 3rd scenario in non-universal $Z'$ model. The maximum value of forward backward asymmetry for $B \to K_2^* \mu\tau$ decay is calculated as $(-0.347)$ with the contribution of vector leptoquark. On the other hand, the average values of $A_{FB}$ are constrained in non-universal $Z'$ model as $(-0.086)$ for 1st scenario, $0.053$ for 2nd scenario and $(-0.025)$ for 3rd scenario. These decays are not studied so far at experiments. The study of these LFV decays in various NP models establishes BSM physics. It can be expected that the new run of LHCb and Belle II experiments will inspect the lepton flavour violation more accurately and will be able to observe the contribution of the NP particles. We hope that the results obtained in Table-2, 3 and 4 will be very useful to the investigation in near future.

## Acknowledgement

We thank the reviewer for constructive comments and suggestion which improve the quality of our paper. S. Biswas and S. Mahata thank NIT Durgapur for providing fellowship for their research. M. Mandal acknowledges the Department of Science and Technology, Govt. of India for providing INSPIRE Fellowship through IF 200277. We would like to thank Dr. Damir Becirevic, Laboratoire de Physique Theorique, CNRS and Univ. Paris-Sud, Universite Paris-Saclay, France, for useful discussions.

## Appendix A

Expressions of transversity amplitudes:

The transversity amplitudes can be represented as

$$A_{L,R}^0 = N \frac{\sqrt{\lambda}}{2\sqrt{6} m_B m_{K_2^*}^2 \sqrt{q^2}} \left[ (C_9^{NP} \mp C_{10}^{NP}) \left\{ (m_B^2 - m_{K_2^*}^2 - q^2)(m_B + m_{K_2^*}) A_1 - \frac{\lambda}{(m_B + m_{K_2^*})} A_2 \right\} \right], \qquad (A1)$$

$$A_{L,R}^\perp = -\sqrt{2} N \frac{\sqrt{\lambda}}{\sqrt{8} m_B m_{K_2^*}} \left[ (C_9^{NP} \mp C_{10}^{NP}) \frac{\sqrt{\lambda} V}{(m_B + m_{K_2^*})} \right], \qquad (A2)$$

$$A_{L,R}^\parallel = \sqrt{2} N \frac{\sqrt{\lambda}}{\sqrt{8} m_B m_{K_2^*}} \left[ (C_9^{NP} \mp C_{10}^{NP})(m_B + m_{K_2^*}) A_1 \right], \qquad (A3)$$

$$A_L^t = N \frac{\lambda}{\sqrt{6} m_B m_{K_2^*} \sqrt{q^2}} \left[ (C_9^{NP} \mp C_{10}^{NP}) A_0 \right], \qquad (A4)$$

where the normalization constant is defined as

$$N = \sqrt{\frac{G_F^2 \alpha_e^2 |V_{tb}V_{ts}^*|^2 q^2}{3 \times 2^{10} m_B^3 \pi^5}} \beta_+ \beta_- \sqrt{\lambda} \mathcal{B}(K_2^* \to K\pi) .$$

The Kallen function can be expressed as $\lambda(m_B^2, m_{K_2^*}^2, q^2) = m_B^4 + m_{K_2^*}^4 + q^4 - 2(m_B^2 m_{K_2^*}^2 + m_B^2 q^2 + m_{K_2^*}^2 q^2)$.

## Appendix B

The $q^2$ dependent form factors are parameterized by the light cone QCD sum rule as

$$F^{B_q T}(q^2) = \frac{\sum_{n=0}^{1} \alpha_n^F \{z(q^2) - z(0)\}^n}{1 - \left(\frac{q^2}{m_{R,F}^2}\right)} . \tag{B1}$$

Where $z(q^2) = \frac{\sqrt{t_+ - s} - \sqrt{t_+ - t_0}}{\sqrt{t_+ - s} + \sqrt{t_+ - t_0}}$, $t_\pm = (m_B + m_{K_2^*})^2$ and $t_0 = t_+(1 - \sqrt{1 - t_-/t_+})$. Here, the term $m_{R,F}$ is known as the resonance mass associated with the corresponding quantum number of the form factor. The fitted values of $\alpha_n^F$ are tabulated below.

| $\alpha_n^F$ | $\alpha_0$ | $\alpha_1$ |
|---|---|---|
| V | $0.22^{+0.11}_{-0.08}$ | $-0.90^{+0.37}_{-0.50}$ |
| $A_0$ | $0.30^{+0.06}_{-0.05}$ | $-1.23^{+0.23}_{-0.23}$ |
| $A_1$ | $0.19^{+0.09}_{-0.07}$ | $-0.46^{+0.19}_{-0.25}$ |
| $A_2$ | $0.11^{+0.05}_{-0.06}$ | $-0.40^{+0.23}_{-0.16}$ |

The uncertainties which arise in the form factors are a result of variation of the input parameters associated with the light cone sum rule (LCSR) calculation. Among the various input parameters, the non-perturbative parameters along with the continuum threshold chiefly contribute to these uncertainties [47].

## Appendix C

The polarization state of $K_2^*$ meson $\epsilon^{\mu\nu}(n)$ can be written as

$$\epsilon_{\mu\nu}(\pm 2) = \epsilon_\mu(\pm 1)\epsilon_\nu(\pm 1),$$

$$\epsilon_{\mu\nu}(\pm 1) = \frac{1}{\sqrt{2}}[\epsilon_\nu(\pm)\epsilon_\nu(0) + \epsilon_\nu(\pm)\epsilon_\mu(0)],$$

$$\epsilon_{\mu\nu}(0) = \frac{1}{\sqrt{6}}[\epsilon_\mu(+)\epsilon_\nu(-) + \epsilon_\nu(+)\epsilon_\mu(-)] + \sqrt{\frac{2}{3}} \epsilon_\mu(0)\epsilon_\nu(0), \tag{C1}$$

where the spin-1 polarization vectors are represented as

$$\epsilon_\mu(0) = \frac{1}{m_{K_2^*}}(\vec{k}_z, 0, 0, k_0), \qquad \epsilon_\mu(\pm) = \frac{1}{\sqrt{2}}(0, 1, \pm i, 0). \tag{C2}$$

Here, in this work the helicity states of $K_2^*$ meson $n = \pm 2$ is not realized and a polarization vector is newly introduced as

$$\epsilon_{T_\mu}(h) = \frac{\epsilon_{\mu\nu} p^\nu}{m_B}.$$

The expressions of polarization vectors are explicitly structured as

$$\epsilon_{T_\mu}(\pm 1) = \frac{1}{m_B} \frac{1}{\sqrt{2}} \epsilon(0).p\epsilon_\mu(\pm) = \frac{\sqrt{\lambda}}{\sqrt{8} m_B m_{K_2^*}} \epsilon_\mu(\pm),$$

$$\epsilon_{T_\mu}(0) = \frac{1}{m_B} \sqrt{\frac{2}{3}} \epsilon(0).p\epsilon_\mu(0) = \frac{\sqrt{\lambda}}{\sqrt{6} m_B m_{K_2^*}} \epsilon_\mu(0). \quad (C3)$$

The term $\lambda(m_B^2, m_{K_2^*}^2, q^2)$ is already defined.

The virtual gauge boson also has three types of polarization states which are longitudinal, transverse and time-like. The components are defined below as

$$\epsilon_V^\mu(0) = \frac{1}{\sqrt{q^2}}(-|\vec{q}_z|, 0,0, -q_0), \epsilon_V^\mu(\pm) = \frac{1}{\sqrt{2}}(0,1,\pm i, 0), \epsilon_V^\mu(t) = \frac{1}{\sqrt{q^2}}(q_0, 0,0, q_z), \quad (C4)$$

where $q^\mu = (q_0, 0,0, q_z)$ is four momentum of gauge boson.

## Appendix D

To study the decay distribution, we need the leptonic matrix elements along with the hadronic matrix elements. Here, we have used the strategy of the ref. [46, 80]. We can define the leptonic matrix elements as follows

$$\langle l_1(\lambda_1) \bar{l}_2(\lambda_2)|\bar{l}\Gamma^X l|0\rangle = \bar{u}(\lambda_1)\Gamma^X v(\lambda_2) = L(\lambda_1, \lambda_2). \quad (D1)$$

Here, the term $\Gamma^X$ is the leptonic parts of the NP operators. The spinors are defined as

$$u_{\frac{1}{2}} = \begin{pmatrix} \sqrt{E_1 + m_1} \\ 0 \\ \sqrt{E_1 - m_1} \\ 0 \end{pmatrix}, u_{-\frac{1}{2}} = \begin{pmatrix} 0 \\ \sqrt{E_1 + m_1} \\ 0 \\ -\sqrt{E_1 - m_1} \end{pmatrix},$$

$$v_{\frac{1}{2}} = \begin{pmatrix} \sqrt{E_2 - m_2} \\ 0 \\ -\sqrt{E_2 + m_2} \\ 0 \end{pmatrix}, v_{-\frac{1}{2}} = \begin{pmatrix} 0 \\ \sqrt{E_2 - m_2} \\ 0 \\ \sqrt{E_2 + m_2} \end{pmatrix}. \quad (D2)$$

We can define the leptonic energy terms as

$E_{1,2} = \sqrt{m_{1,2}^2 + \lambda(m_1^2, m_2^2, q^2)/4q^2}$ and $E_1 + E_2 = \sqrt{q^2}$. The normalized spinors are:

$$\bar{u}(\lambda_1)u(\lambda_2) = \delta_{\lambda_1 \lambda_2} 2m_1, \qquad \bar{v}(\lambda_1)v(\lambda_2) = -\delta_{\lambda_1 \lambda_2} 2m_2$$

Along with all the expressions we have structured the leptonic helicity amplitudes as below.

$$L^{L,R}(1/2, 1/2, 0) = (m_- \beta_+ \pm m_+ \beta_-)/2,$$
$$L^{L,R}(-1/2, -1/2, 0) = (m_- \beta_+ \mp m_+ \beta_-)/2,$$
$$L^{L,R}(1/2, 1/2, 1) = (m_+ \beta_- \pm m_- \beta_+)/2,$$
$$L^{L,R}(-1/2, -1/2, 1) = (m_+ \beta_- \mp m_- \beta_+)/2,$$
$$L^{L,R}(-1/2, 1/2, 1) = -\sqrt{q^2}(\beta_- \pm \beta_+)/\sqrt{2},$$
$$L^{L,R}(1/2, -1/2, 1) = -\sqrt{q^2}(\beta_+ \pm \beta_-)/\sqrt{2}. \quad (D3)$$

Here, $\beta_\pm = \sqrt{1 - \frac{(m_1 \pm m_2)^2}{q^2}}$ and $m_\pm = (m_1 \pm m_2)$. Other than these above written helicity amplitudes are zero.

**Appendix E:**

**Table-8: Values of other input parameters [72]**

| Parameter | Values |
|---|---|
| $m_\mu$ | $(105.66 \pm 0.0000024)$ MeV |
| $m_e$ | $(0.51 \pm 0.0000000031)$ MeV |
| $m_\tau$ | $(1776.86 \pm 0.12)$ MeV |
| $m_{K_2^*}$ | $(1427.3 \pm 1.5)$ MeV |
| $m_B$ | $(5279.55 \pm 0.26)$ MeV |
| $G_F$ | $(1.166 \pm 0.0000006) \times 10^{-5}$ GeV$^{-2}$ |
| $|V_{tb}|$ | $(1.019 \pm 0.025)$ |
| $|V_{ts}|$ | $(39.4 \pm 2.3) \times 10^{-3}$ |

**References**

1. M. Huschle et al. [Belle Collaboration], "Measurement of the branching ratio of $\bar{B} \to D^{(*)}\tau^-\bar{\nu}_\tau$ relative to $\bar{B} \to D^{(*)}l^-\bar{\nu}_l$ decays with hadronic tagging at Belle", *Phys. Rev. D* **92**, 072014 (2015) [arXiv: 1507.03233[hep-ex]].
2. S. Hirose et al. [Belle Collaboration], "Measurement of the $\tau$ lepton polarization and $R(D^*)$ in the decay $\bar{B} \to D^*\tau^-\bar{\nu}_\tau$", *Phys. Rev. Lett.* **118**, 211801 (2017) [arXiv: 1612.00529 [hep-ex]].
3. J. P. Lees et al. [BaBar Collaboration], "Measurement of an excess of $\bar{B} \to D^*\tau^-\bar{\nu}_\tau$ decays and implications for charged higgs bosons", *Phys. Rev. D* **88**, 072012 (2013) [arXiv: 1303.0571[hep-ex]].
4. G. Ciezarek, on behalf of the LHCb collaboration, CERN, "$R(D^*)$ and $R(D)$ with $\tau^- \to \mu^-\nu_\tau\bar{\nu}_\mu$". http://indico.cern.ch/event/1187939.
5. G. Caria et al. [Belle Collaboration], "Measurement of $R(D)$ and $R(D^*)$ with a semileptonic tagging method", *Phys. Rev. Lett.* **124**, 161803 (2020) [arXiv: 1910.05864 [hep-ex]].
6. D. Bigi and P. Gambino, "Revisiting $B \to Dl\nu$", *Phys. Rev. D* **94**, 094008 (2016) [arXiv: 1606.08030 [hep-ph]].

7. S. Jaiswal, S. Nandi and S. K. Patra, "Extracting of $|V_{cb}|$ from $B \to D^{(*)} l\nu_l$ and the Standard model predictions of $R(D^{(*)})$", *JHEP* **1712**, 060 (2017) [arXiv: 1707.09977 [hep-ph]].
8. HFLAV Collaboration, "Average of $R(D)$ and $R(D^*)$ for end of 2022", http://hflav.eos.web.cern.ch/hflav-eos/semi/fall22/html/RDsDsstar/RDRDs.html.
9. LHCb Collaboration, "Test of lepton universality in $b \to sl^+l^-$ decays" [arXiv: 2212.09152 [hep-ex]].
10. LHCb Collaboration, "Measurement of lepton universality parameters in $B^+ \to K^+l^+l^-$ and $B^0 \to K^{*0}l^+l^-$ decays" [arXiv: 2212.09153 [hep-ph]].
11. R. Aaij et al. [LHCb Collaboration], "Search for lepton-universality violation in $B^+ \to K^+l^+l^-$ decays" *Phys. Rev. Lett.* **122**, 191801 (2019) [arXiv: 1903.09252 [hep-ex]]; "Test of lepton universality with $B^0 \to K^{*0}l^+l^-$ decays", *JHEP* **1708**, 055 (2017) [arXiv: 1705.05802 [hep-ex]].
12. A. Abdesselam et al. [Belle Collaboration], "Test of lepton flavor universality in $B \to K^*l^+l^-$ decays at Belle", [arXiv: 1904.02440 [hep-ex]].
13. L. Calibbi and G. Signorelli, "Charged lepton flavour violation: An experimental and theoretical introduction", *Riv. Nuovo Cim.* **41,** 71 (2018) [arXiv: 1709.00294 [hep-ph]].
14. R. Aaij et al. [LHCb Collaboration], "Search for the lepton-flavour-violating decays $B_s^0 \to \tau^{\pm}\mu^{\mp}$ and $B^0 \to \tau^{\pm}\mu^{\mp}$", *Phys. Rev. Lett.* **123**, 211801 (2019) [arXiv: 1905.06614 [hep-ex]].
15. L. Nayak et al. [Belle Collaboration], "Search for $B_s^0 \to l^{\pm}\mu^{\mp}$ with the semi-leptonic tagging method at Belle," arXiv: 2301.10989 [hep-ex].
16. T. M. Aliev and M. Savel, "$B \to K_2 l^+l^-$ decay beyond the standard model", *Phys. Rev. D* **85**, 015007 (2012) [arXiv: 1109.2738 [hep-ph]].
17. H. Hatanaka and K. C. Yang, "Radiative and Semileptonic $B$ Decays Involving the Tensor Meson $K_2^*(1430)$ in the Standard Model and Beyond", *Phys. Rev. D* **79**, 114008 (2009) [arXiv: 0903.1917 [hep-ph]].
18. R. Mohanta, "Effect of FCNC mediated Z boson on lepton flavour violating decays", *Eur. Phys. J. C* **71**, 1625 (2011) [arXiv: 1011.4184 [hep-ph]].
19. P. Nayek, S. Biswas, P. Maji and S. Sahoo, "Implication of Z-mediated FCNC on semileptonic decays $B_s \to \varphi l^+l^-$ and $B^+ \to K^+l^+l^-$", *Int. J. Theor. Phys.* **59,** 1418 (2020).
20. S. Sahoo and R. Mohanta, "New physics effects in charm meson decays involving $c \to u l^+l^- (l_i^{\mp} l_j^{\pm})$ transitions", *Eur. Phys. J. C* **77**, 344 (2017) [arXiv: 1705.02251 [hep-ph]].
21. A. Crivellin et al., "Lepton-flavour violating B decays in generic $Z'$ models", *Phys. Rev. D* **92**, 054013 (2015) [arXiv: 1504.07928 [hep-ph]].
22. Y. Farzan and I. M. Shoemaker, "Lepton flavour violating non-standard interactions via light mediators", *JHEP* **1607**, 033 (2016) [arXiv: 1512.09147 [hep-ph]].
23. S. Sahoo and R. Mohanta, "Lepton flavour violating $B$ meson decays via scalar leptoquark", *Phys. Rev. D* **93,** 114001 (2016) [arXiv:1512.04657[hep-ph]].
24. M. Duraisamy, S. Sahoo and R. Mohanta, "Rare semileptonic $B \to K(\pi)l_i^- l_j^+$ decay in vector leptoquark model", *Phys. Rev. D* **95**, 035022 (2017) [arXiv: 1610.00902 [hep-ph]].

25. S. Sahoo and R. Mohanta, "Effects of scalar leptoquark on semileptonic $\Lambda_b$ decays", *New J. Phys.* **18**, 093051 (2016) [arXiv: 1607.04449 [hep-ph]].
26. D. Das, "Lepton flavour violating $\Lambda_b \to \Lambda l_1 l_2$ decay", *Eur. Phys. J. C* **79**, 1005 (2019) [arXiv: 1909.08676 [hep-ph]].
27. A. Dedes, J. Rosiek and P. Tanedo, "Complete one-loop MSSM predictions for $B^0 \to l^+ l^{-\prime}$ at the Tevatron and LHC", *Phys. Rev. D* **79**, 055006 (2009) [arXiv: 0812.4320 [hep-ph]].
28. U. Chattopadhyay, D. Das and S. Mukherjee, "Probing lepton flavour violating dcays in MSSM with non-holomorphic soft terms", *JHEP* **2006**, 015 (2020) [arXiv: 1911.05543 [hep-ph]].
29. A. Crivellin et al., "Lepton flavour violation in the MSSM: exact diagonalization vs mass expansion", *JHEP* **1806**, 003 (2018) [arXiv: 1802.06803 [hep-ph]].
30. D. Becirevic, O. Sumensari and R. Zukanovich Funchal, "Lepton flavour violation in exclusive $b \to s$ decays", *Eur. Phys. J. C* **76**, 134 (2016), [arxiv; 1602.00881 [hep-ph]].
31. B. M. Dassinger, Th. Feldmann, Th. Mannel and S. Turczyk, "Model-independent analysis of lepton flavour violating tau decays", *JHEP* **0710**, 039 (2007) [arXiv: 0707.0988 [hep-ph]].
32. S. Biswas, P. Nayek, P. Maji and S. Sahoo, "Lepton flavour violating $\Lambda_b$ decays in non-universal $Z'$ model", *Eur. Phys. J. C* **81**, 493 (2021).
33. S. Biswas, S. Mahata, A. Biswas and S. Sahoo, "Impact of non-universal $Z'$ in the lepton flavour violating $B(B_s) \to K^*(\varphi) l_1^- l_2^+$ decays", *Eur. Phys. J. C* **82**, 578 (2022).
34. F. Abe et al. [CDF Collaboration], "Inclusive jet cross section in $\bar{p}p$ collisions at $s\} = 1.8$ TeV", *Phys. Rev. Lett.* **77**, 438 (1996).
35. F. Abe et al. [CDF Collaboration], "Measurement of the $e^+e^-$ invariant-mass distribution in $p\bar{p}$ collisions at $\sqrt{s} = 1.8$ TeV", *Phys. Rev. Lett.* **67**, 2418 (1991).
36. D. Abbaneo et al. [LEP Collaborations], "A Combination of preliminary electroweak measurements and constraints on the Standard Model", [arXiv: hep-ex/0212036].
37. S. Sahoo, C. K. Das and L. Maharana, "The prediction of mass of $Z'$-boson from $B_q^0 - \overline{B_q^0}$ Mixing", *Int. J. Mod. Phys. A* **26**, 3347 (2011) [arXiv: 1112.0460 [hep-ph]].
38. M. Aaboud et al. [ATLAS Collaboration], "Search for additional heavy neutral Higgs and gauge bosons in the ditau final state produced in 36 $fb^{-1}$ of pp collisions at $\sqrt{s} = 13$ TeV with the ATLAS detector", *JHEP* **1801**, 055 (2018) [arXiv: 1709.07242 [hep-ex]].
39. A. M. Sirunyan et al. [CMS Collaboration], "Search for high-mass resonances in dilepton final states in proton-proton collisions at $\sqrt{s} = 13$ TeV", *JHEP* **1806**, 120 (2018) [arXiv: 1803.06292 [hep-ex]].
40. T. Bandyopadhyay, G. Bhattacharya, D. Das and A. Raychaudhuri, "A reappraisal of constraints on $Z'$ models from unitarity and direct searches at the LHC", *Phys. Rev. D* **98**, 035027 (2018) [arXiv: 1803.07989 [hep-ph]].
41. M. Aaboud et al. [ATLAS Collaboration], "Search for new high-mass phenomena in the dilepton final state using 36 $fb^{-1}$ of proton-proton collision data at $\sqrt{s} = 13$ TeV with the ATLAS detector", *JHEP* **1710**, 182 (2017).

42. G. Aad et al. [ATLAS Collaboration], "Search for high-mass dilepton resonances using 139 $fb^{-1}$ of $pp$ collision data collected at $\sqrt{s} = 13$ TeV with the ATLAS detector", *Phys. Lett. B* **796**, 68 (2019).
43. L. D. Luzio, M. Kirk, A. Lenz and T. Rauh, "$\Delta M_s$ theory precision confronts flavour anomalies", *JHEP* **1912**, 009 (2019) [arXiv: 1909.11087 [hep-ph]].
44. Q. Chang, X-Qiang Li and Y-Dong, "Family non-universal $Z'$ effects on $\overline{B_q} - B_q$ mixing, $B \to X_s \mu^+\mu^-$ and $B_s \to \mu^+\mu^-$ Decays", *JHEP* **1002**, 082 (2010) [arXiv: 0907.4408 [hep-ph]].
45. S. Kumbhakar, R. Sain and J. Vardani, "Lepton flavour violating $B \to K_2^*(1430)\mu^\pm\tau^\mp$ decays", arXiv: 2208.05923 [hep-ph].
46. D. Das, "Lepton flavour violating $\Lambda_b \to \Lambda l_1 l_2$ decay", *Eur. Phys. J. C* **79**, 1005 (2019) [arXiv: 1909.08676 [hep-ph]].
47. T. M. Aliev, H. Dag, A. Kokulu and A. Ozpineci, "$B \to T$ transition form factors in light-cone sum rules", *Phys. Rev. D* **100**, 094005 (2019) [arXiv: 1908.00847 [hep-ph]].
48. P. Langacker and M. Plumacher, "Flavor changing effects in theories with a heavy $Z'$ boson with family non universal couplings", *Phys. Rev. D* **62**, 013006 (2000) [arXiv: hep-ph/0001204].
49. V. Barger et al., "$b \to s$ transitions in family-dependent $U(1)'$ models", *JHEP* **0912**, 048 (2009).
50. S. Chaudhuri, S. W. Chung, G. Hockney and J. Lykken, "String consistency for unified model building", *Nucl. Phys. B* **456**, 89 (1995) [arXiv: hep-ph/9501361].
51. G. Cleaver, M. Cvetič, J. R. Epinosa, L. Everett and P. Langacker, "Classification of flat directions in perturbative heterotic superstring vacua with anomalous $U(1)$", *Nucl. Phys. B* **525**, 3 (1998).
52. G. Cleaver, M. Cvetič, J. R. Espinosa, L. Everett, P. Langacker and J. Wang, "Physics implications of flat directions in free fermionic superstring models. I: Mass spectrum and couplings", *Phys. Rev. D* **59**, 055005 (1999).
53. P. Abreu et al. [DELPHI Collaboration], "A study of radiative muon-pair events at $Z^0$ energies and limits on an additional $Z'$ gauge boson", *Z. Phys. C* **65**, 603 (1995).
54. J. Erler and P. Langacker, "Constraints on extended neutral gauge structures", *Phys. Lett. B* **456**, 68 (1999) [arXiv: hep-ph/9903476].
55. J. Erler and P. Langacker, "Indications for an extra neutral gauge boson in electroweak precision data", *Phys. Rev. Lett.* **84**, 212 (2000) [arXiv: hep-ph/9910315].
56. J. Erler, P. Langacker, S. Munir and E. R. Pena, "Improved constraints on $Z'$ bosons from electroweak precision data", *JHEP* **0908**, 017 (2009) [arXiv: 0906.2435 [hep-ph]].
57. I. D. Bobovnikov, P. Osland and A. A. Pankov, "Improved constraints on the mixing and mass of $Z'$ bosons from resonant diboson searches at the LHC at $\sqrt{s} = 13$ TeV and predictions for Run II", *Phys. Rev. D* **98**, 095029 (2018) [arXiv: 1809.08933 [hep-ph]].
58. V. Barger, C-W. Chiang, P. Langacker and H.S. Lee, "Solution to the $B \to \pi K$ Puzzle in a Flavor-Changing $Z'$ Model", *Phys. Lett. B* **598**, 218 (2004) [arXiv: hep-ph/0406126].

59. V. Barger, C-W. Chiang, P. Langacker and H.S. Lee, "$Z'$ mediated flavour changing neutral currents in $B$ meson decays", *Phys. Lett. B* **580**, 186 (2004) [arXiv: hep-ph/0310073].

60. V. Barger, C-W. Chiang, J. Jiang and P. Langacker, "$B_s - \bar{B}_s$ mixing in $Z'$ models with flavor-changing neutral currents", *Phys. Lett. B* **596**, 229 (2004) [arXiv: hep-ph/0405108].

61. R. Mohanta and A. K. Giri, "Explaining $B \to \pi K$ anomaly with non-universal $Z'$ boson", *Phys. Rev. D* **79**, 057902 (2009) [arXiv: 0812.1842 [hep-ph]].

62. K. Cheung, C-W. Chiang, N. G. Deshpande and J. Jiang, "Constraints on flavor-changing $Z'$ models by $B_s$ mixing, $Z'$ production, and $B_s \to \mu^+\mu^-$", *Phys. Lett. B* **652**, 285 (2007) [arXiv: hep-ph/0604223].

63. M. Bona et al. [UTfit Collaboration], First evidence of new physics in $b \leftrightarrow s$ transitions, *PMC Phys. A* **3,** 6 (2009) [arXiv: 0803.0659 [hep-ph]].

64. M. Bona et al. [UTfit Collaboration], "New physics from flavour", [arXiv: 0906.0953 [hep-ph]].

65. C. H. Chen, "CP phase of nonuniversal $Z'$ on $\sin \phi_s^{J/\phi}$ and $T$-odd observables of $\bar{B}_q \to V_q l^+ l^-$", *Phys. Lett. B*, **683**, 160 (2010) [arXiv: 0911.3479 [hep-ph]].

66. N. G. Deshpande, X. G. He and G. Valencia, "D0 dimuon asymmetry in $B_s - \bar{B}_s$ mixing and constraints on new physics", *Phys. Rev. D* **82,** 056013 (2010) [arXiv: 1006.1682 [hep-ph]].

67. J. E. Kim, M. S. Seo and S. Shin, "D0 same-charge dimuon asymmetry and possible new CP violation sources in the $B_s - \bar{B}_s$ system", *Phys. Rev. D* **83,** 036003 (2011) [arXiv: 1010.5123 [hep-ph]].

68. P. J. Fox, J. Liu, D. Tucker-Smith and N. Weiner, "An effective $Z'$", *Phys. Rev. D* **84**, 115006 (2011) [arXiv: 1104.4127 [hep-ph]].

69. Q. Chang, R. M. Wang, Y. G. Xu and X. W. Cui, "Large dimuon asymmetry and non-universal $Z'$ boson in the $B_s - \bar{B}_s$ system", *Chin. Phys. Lett.* **28,** 081301 (2011).

70. D. Banerjee and S. Sahoo, "Analysis of $\Lambda_b \to \Lambda l^+ l^-$ in a non-universal $Z'$ model", *Chin. Phys. C* **41**, 083101 (2017).

71. S. Biswas, P. Nayek, P. Maji and S. Sahoo, "New Physics Effect on $B_{s,d} \to l^+ l^-$ decays", *Int. J. Theor. Phys.* **60**, 893 (2021).

72. R. L. Workman et al. [Particle Data Group], *Prog. Theor. Exp. Phys.* **2022**, 083C01 (2022).

73. D. Ghosh, M. Nardecchia and S. A. Renner, "Hint of lepton flavour non-universality in $B$ meson decays", *JHEP* **1412**, 131 (2014) [arXiv: 1408.4097 [hep-ph]].

74. D. Becirevic, N. Kosnik, O. Sumensari and R. Z. Funchal, "Palatable leptoquark scenarios for lepton flavour violation in exclusive $b \to s l_1 l_2$ modes", *JHEP* **1611**, 035 (2016) [arXiv: 1608.07583 [hep-ph]].

75. D. Das, "On the angular distribution $\Lambda_b \to \Lambda(\to N\pi)\tau^+\tau^-$ decay", *JHEP* **1807**, 063 (2018) [arXiv: 1804.08527[hep-ph]].

76. A. Datta, J. Kumar and D. London, "The $B$ Anomalies and new physics in $b \to s e^+ e^-$", *Phys. Lett. B* **797**, 134858 (2019) [arXiv: 1903.10086 [hep-ph]].

77. A. K. Alok, A. Dighe, S. Gangal and D. Kumar, "Continuing search for new physics in $b \to s\mu\mu$ decays: two operators at a time", *JHEP* **1906**, 089 (2019) [arXiv: 1903.09617 [hep-ph]].
78. G. Hiller and M. Schmatz, "$R_K$ and future $b \to sll$ BSM opportunities", *Phys. Rev. D* **90**, 054014 (2014) [arXiv: 1408.1627 [hep-ph]].
79. D. Das, B. Kindra, G. Kumar and N. Mahajan, "$B \to K_2^*(1430)l^+l^-$ distributions in standard model at large recoil", *Phys. Rev. D* **99**, 093012 (2019) [arXiv: 1812.11803 [hep-ph]].
80. H. E. Haber, "Spin formalism and applications to new physics searches", In *Stanford 1993, Spin structure in high energy processes* 231-272 [arXiv: hep-ph/9405376].